\documentclass[a4paper]{article}

\usepackage{INTERSPEECH2021}
\usepackage{dblfloatfix} 
\usepackage{array}
\newcolumntype{P}[1]{>{\centering\arraybackslash}p{#1}}

\newcommand{\Ac}{\mathcal{A}}

\newcommand{\Lc}{\mathcal{L}}

\newcommand{\Pc}{\mathcal{P}}

\newcommand{\Tc}{\mathcal{T}}

\usepackage{multirow}
\usepackage{url}            

\title{Cellular Network Speech Enhancement: Removing Background and Transmission Noise}
\name{Amanda Shu, Hamza Khalid, Haohui Liu, Shikhar Agnihotri, Joseph Konan, Ojas Bhargave}
\address{
  Carnegie Mellon University}
\email{\texttt{\{amshu, hkhalid, haohuil, sagnihot, jkonan, obhargav\}@andrew.cmu.edu}}

\begin{document}

\maketitle
 
 
\begin{figure*}[h]
  \centering
  \includegraphics[width=\linewidth,height=5cm]{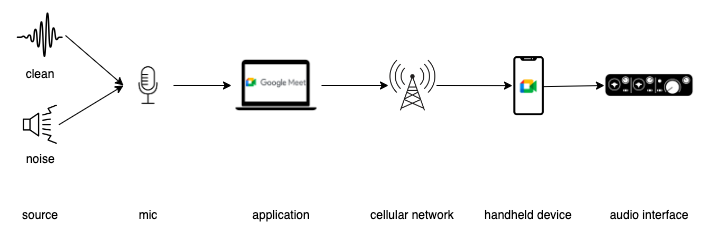}
  \caption{End-to-end Data Process Diagram, From Synthesis To Recording. }
  \label{fig:transmission_diagram}
\end{figure*}

\begin{abstract}
  The primary objective of speech enhancement is to reduce background noise while preserving the target's speech. A common dilemma occurs when a speaker is confined to a noisy environment and receives a call with high background and transmission noise. To address this problem, the Deep Noise Suppression (DNS) Challenge focuses on removing the background noise with the next-generation deep learning models to enhance the target’s speech; however, researchers fail to consider Voice Over IP (VoIP) applications their transmission noise. Focusing on Google Meet and its cellular application, our work achieves state-of-the-art performance on the Google Meet To Phone Track of the VoIP DNS Challenge. This paper demonstrates how to beat industrial performance and achieve 1.92 PESQ and 0.88 STOI, as well as superior acoustic fidelity, perceptual quality, and intelligibility in various metrics.
\end{abstract}

\section{Introduction}

Online Voice Over IP (VoIP) meeting platforms such as Google Meet, Zoom, and Microsoft Teams allow users to join conferences through a regular phone via Audio Conferencing \cite{ilag2020teams}. This option is frequently used by users who are either on the go, have a poor internet connection, or would prefer to join a meeting hands-free \cite{ilag2022microsoft}. Given the convenience such an option offers, it is an indispensable functionality that motivates our focus on the Google Meet To Phone track of the VoIP DNS Challenge. \footnote{\url{https://github.com/deepology/VoIP-DNS-Challenge}} 

However, Audio Conferencing brings about the problem of transmission noise. Speech quality is degraded not only by the background noise present when taking the call but also by the transmission channel, whether it be through a telephone line or via wireless transmission \cite{kaiser2018impact}. This is even more prevalent with mobile telephone transmission, where additional factors, including network congestion and packet loss, further degrade transmission quality \cite{mcdougall2015telephone}. These factors result in information loss during communication \cite{lawrence2008acoustic}.

Thus, it is important to mitigate information loss by introducing speech enhancement methods to reduce transmission noise. Existing deep learning approaches predominantly focus on removing background noise, and Google Meet itself has a “Noise Cancellation” feature that uses deep learning models to address this issue. However, speech enhancement to remove transmission noise and background noise remains largely unexplored. Our aim is to explore speech enhancement models and improve the speech quality of audio coming through Google Meets over calls. The code for all the experiments and ablations can be found at \url{ https://github.com/hamzakhalidhk/11785-project}

\section{Background}

Within the current literature, datasets for speech enhancement tasks are often synthesized, a process in which noise is added to clean audio. This is because supervised optimization requires pairs of noisy inputs and clean targets.
\cite{reddy2020interspeech} introduced the INTERSPEECH 2020 Deep Noise Suppression Challenge (DNS) with a reproducible synthesis methodology. With clean speech from Librivox corpus and noise from Freesound and AudioSet, enabling upwards of 500 hours of noisy speech to be created. The VoIP DNS Challenge provides 20 hours of Google Meets to Phone relay, including synthetic background noise at the source and real-world transmission noise at the receiver. 

Of the current state-of-the-art speech enhancement models on the 2020 DNS Challenge, this study focuses on Demucs and FullSubNet. Demucs operates on waveform inputs in the time domain, whose real-time denoising architecture was proposed by Defossez et al. \cite{defossez2019music}. FullSubNet operates on spectrogram inputs in the time-frequency domain and was developed by Hao et al. \cite{hao2021fullsubnet}. Baseline Demucs achieves 
1.73 PESQ and 0.86 STOI, 
while Baseline FullSubNet achieves
1.69 PESQ and 0.84 STOI.
As both of these baseline models do not account for transmission noise, our investigation is focused on fine-tuning and improving the acoustic fidelity, perceptual quality, and intelligibility for the VoIP DNS Challenge challenge.


%

\section{Dataset}
\label{headings}

Our dataset is built off of audio clips from the open-sourced dataset\footnote{\url{https://github.com/microsoft/DNS-Challenge/tree/interspeech2020/master/datasets}} released from Microsoft's Deep Noise Suppression Challenge (DNS). Specifically, we utilize the test set of synthetic clips without reverb from the DNS data, as described by Reddy et al. \cite{reddy2020interspeech}. Our novel dataset contains these audio clips with the addition of transmission noise resulting from taking a Google Meet call through a cellular device. It is created as described in the following steps.

First, clean speech and noise audio are randomly selected from the DNS dataset. These audio files are then mixed, resulting in the noisy audio of clean speech with background noise. Next, to include the transmission noise of a Google Meet to a phone session, we start a Google Meet session and call into the session using a phone that is on the T-mobile network. We then play the noisy speech audio in Google Meet, and the audio that is relayed to the phone is recorded to an audio interface.

The final recorded audio captures both background noise and transmission noise on a Google Meet session. Furthermore, we note that we have recorded the audio with both the Google Meet noise cancellation feature turned on and off to compare our models with industry standards. We refer to the audio with Google Meet speech enhancement turned off as \textit{low}, and the audio with speech enhancement turned on as \textit{auto}. In our work, for each of the auto and low data sources, we utilize 400 audio clips (each thirty seconds long) for our training samples and 150 ten-second audio clips for our testing samples. We further apply gain normalization to transform the audio files to the same amplitude range to facilitate training. This transforms the audio files to the same amplitude range to enable better model training. Figure~\ref{fig:transmission_diagram} depicts the dataset synthesis process.

\begin{figure*}[t]
  \centering
  \includegraphics[width=\linewidth]{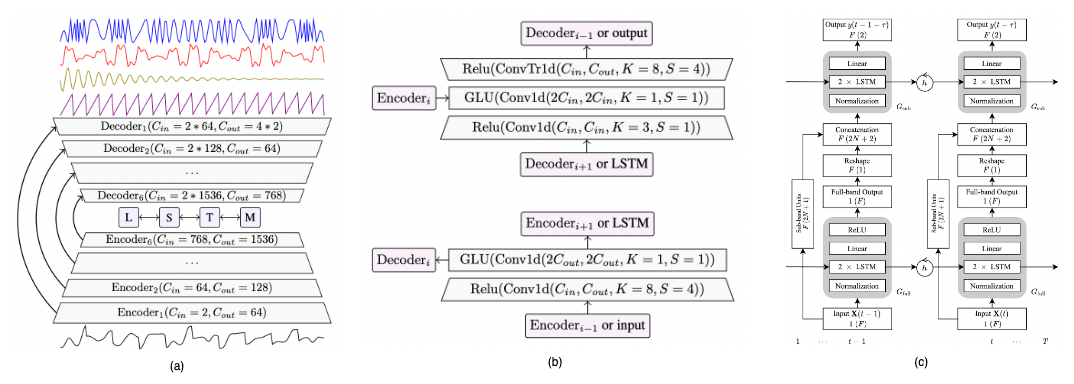}
  \caption{(a) Demucs architecture with the mixture waveform as input and the four sources estimates
as output. Arrows represent U-Net connections. (b) Detailed view of the layers $Decoder_i$ on the top
and $Encoder_i$ on the bottom. Arrows represent connections to other parts of the model. (c) FullSubNet architecture. The second line in the rectangle describes the dimensions of the data at the current stage, e.g., “1 (F)” represents one F-dimensional vector. “F (2N + 1)” represents F independent (2N + 1)-dimensional vectors.}
  \label{fig:arch}
\end{figure*}

\section{Models}


We select Demucs and FullSubNet  (see Section \ref{appendix-demucs} and \ref{appendix-fsn}), as baseline models in our work, as both models have yielded state-of-the-art results in the DNS Challenge \cite{hao2021fullsubnet, defossez2020real}. Demucs consists of 2 LSTM layers between an encoder-decoder structure. FullSubNet is a fusion model that combines a full-band model that captures the global spectral context and a sub-band model that encapsulates the local spectral pattern for single-channel real-time speech enhancement.

\subsection{Demucs}
\label{appendix-demucs}
The Demucs architecture is heavily inspired by the architectures of SING \cite{defossez2018sing}, and Wave-U-Net \cite{stoller2018wave}.
It is composed of a convolutional encoder, an LSTM, and a convolutional decoder. The encoder and decoder are linked with skip U-Net connections. The input to the model is a stereo mixture $s = \sum_i s_i$ and the output is stereo estimate $\hat{s_i}$ for each source. Fig \ref{fig:arch} (a) shows the architecture of the complete model.

\label{demuc_loss}
Demucs' criterion minimizes the sum of the L1-norm, $\Lc_1$, between waveforms and the multi-resolution STFT loss, $\Lc_{STFT}$ of the magnitude spectrograms.

\begin{align*}
    \Lc_{demucs} 
    & = \frac{1}{T} [||y-\hat{y}||_1 + \sum_{i=1}^{M}\Lc_{STFT}^{(i)}(y,\hat{y})] \\
    & \equiv \Lc_{L1} + \Lc_{STFT}
\end{align*}
Without $\Lc_{STFT}$, we observe tonal artifacts. We discuss ablating $\Lc_{STFT}$ and more recent auxiliary losses in Section 5.2.

\subsection{FullSubNet}
\label{appendix-fsn}
FullSubNet is a full-band and sub-band fusion model, each with a similar topology. This includes two stacked unidirectional LSTM layers and one linear (fully connected) layer. The only difference between the two is that, unlike the full-band model, the output layer of the sub-band model does not use any activation functions. Fig \ref{fig:arch} (c) shows the complete model architecture.

FullSubNet adopts the complex Ideal Ratio Mask ($cIRM$) as their model’s learning target. They use a hyperbolic tangent to compress $cIRM$ in training and an inverse function to uncompress the mask in inference $(K = 10, C = 0.1)$.

\section{Methods}

\begin{figure}[t]
  \centering
  \includegraphics[width=\linewidth]{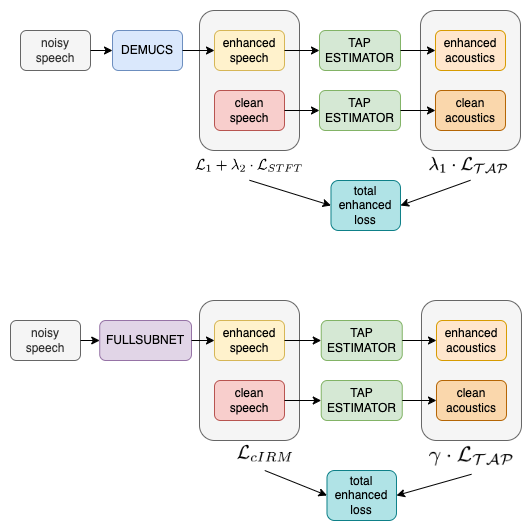}
  \caption{The training workflow for Demucs and FullSubNet}
  \label{fig:train1}
\end{figure}

\subsection{Baseline Method}
On the VoIP DNS Challenge, Google Meets To Phone track, we determine baseline performance using pre-trained Demucs and FullSubNet. Our work improves on this, ablating a variety of criteria. Table 1 and Table 2 of Section \ref{eval-metrics} show ablations. 

\subsection{TAPLoss}
We introduce TAPLoss during training to outperform the state-of-the-art speech recognition models on our data. Taploss involves a set of 25 temporal acoustic parameters, including frequency-related parameters: pitch, jitter, F1, F2, F3 Frequency and bandwidth; energy or amplitude-related parameters: shimmer, loudness, harmonics-to-noise (HNR) ratio; spectral balance parameters: alpha ratio, Hammarberg Index, spectral slope, F1, F2, F3 relative energy, harmonic difference; and additional temporal parameters: rate of loudness peaks, mean and standard deviation of length of voiced/unvoiced regions, and continuous voiced regions per second.

\subsubsection{Enhanced Demucs loss}
Previous works have shown that Demucs model is prone to generating tonal artifacts. The multi-resolution STFT loss amplifies this issue because the error introduced by tonal artifacts is more significant and obvious in the time-frequency domain than in the time domain. Thus, we reduce the influence of $\Lc{STFT}$ in the demucs loss and, additionally, introduce the tap loss $\Lc_{\Tc\Ac\Pc}$. The new demucs loss function is defined as: 
$$\Lc_{Demucs} = \Lc_{1} + \lambda_1 \cdot \Lc_{\Tc\Ac\Pc} + \lambda_2 \cdot \Lc_{STFT}$$

\subsubsection{Enhanced FullSubNet loss}

We also extend the FullSubNet loss by introducing the $\Lc_{\Tc\Ac\Pc}$ loss. The new FullSubNet loss is defined as:
 
$$\Lc_{FullSubNet} = \Lc_{cIRM} + \gamma \cdot \Lc_{\Tc\Ac\Pc}$$

\subsection{Ablations}
We perform ablations to find the optimal values for each of the hyperparameters $\lambda_1$ and $\lambda_2$ for Demucs, and $\gamma$ for FullSubNet. These optimal values were determined by their performance on the Perceptual Evaluation of Speech Quality (PESQ) and Short-Time Objective Intelligibility (STOI) metrics, which are defined in Table ~\ref{table:metrics-table}. Table ~\ref{table:demucs-ablations} and Table ~\ref{table:fsn-ablations} show the summary of our ablations for Demucs and FullSubNet, respectively.

\subsection{Evaluation Metrics}
\label{eval-metrics}
After determining the optimal hyperparameter values for both models, we further evaluate the respective models with both objective metrics and acoustic parameters.

\subsubsection{Objective Evaluation}
 In addition to PESQ and STOI, we further test the models on three more objective metrics: Log-Likelihood Ratio (LLR), Coherence and Speech Intelligibility Index (CSII), and Normalized-Covariance Measure (NCM). PESQ and LLR measure Speech Quality (SQ), while the STOI, CSII, and NCM measure Speech Intelligibility (SI). All four metrics aim to capture the human-judged quality of speech recordings, which is regarded as the gold standard for evaluating Speech Enhancement models \cite{reddy2021dnsmos}. These metrics are defined in Appendix ~\ref{appendix-metrics}.

    
    
    

\subsubsection{Acoustic Evaluation}
In addition to objective metrics, we also utilize the set of acoustic parameters, specifically the eGeMAPSv02 functional descriptors, presented by \cite{eyben2015geneva} to evaluate the model further. This is a set of 88 frequency, energy/amplitude, and spectral-related parameters. We use the OpenSMILE (Open-source Speech and Music Interpretation by Large-space Extraction)\footnote{\url{https://github.com/audeering/opensmile-python}} python package for these parameters. 

\begin{table}
  \caption{Demucs Ablations}
  \label{table:demucs-ablations}
  \centering
    \begin{tabular}{lrrrrrr}
    \toprule[1.5pt] \toprule[1.5pt]
    \multicolumn{7}{c}{Industry Speech Enhancement ON} \\ 
    \toprule[1.2pt]
    $\lambda_1$ &  1.0 &  1.0 &  1.0 &  0.8 & 0.75 &  0.5 \\
    $\lambda_2$ & 0.8 & 0.5 & 0.0 &  0.5 & 0.5 & 0.5 \\
    \midrule
    STOI    &  0.832 & 0.852 &  0.681 &  0.867 &  \textbf{0.882} &  0.870 \\
    PESQ   &  1.658 &  1.844  &  1.488 &  1.780 &  \textbf{1.921} &  1.863 \\
    \bottomrule[1.5pt]
    \end{tabular}
\end{table}

\begin{table}
  \caption{FullSubNet Ablations}
  \label{table:fsn-ablations}
  \centering
    \begin{tabular}{lrrrrrr}
    \toprule[1.5pt] \toprule[1.5pt]
    \multicolumn{7}{c}{Industry Speech Enhancement ON} \\ 
    \toprule[1.2pt]
    $\gamma$ &  1.00 &  0.30 &  0.10 &  0.03 &  0.01 &  0.00 \\
    \midrule
    STOI            &  \textbf{0.864} &  0.863 &  0.860 &  0.861 & 0.858 &  0.861 \\
    PESQ                 &  1.843 &  1.829 &  1.805 &  \textbf{1.867} & 1.787 &  1.808 \\
    \toprule[1.4pt] \toprule[1.4pt]
    \multicolumn{7}{c}{Industry Speech Enhancement OFF} \\ 
    \toprule[1.2pt]
    $\gamma$ &  1.00 &  0.30 &  0.10 &  0.03 &  0.01 &  0.00 \\
    \midrule
    STOI            &  0.722 &  0.731 &  0.735 & \textbf{0.740} &  0.739 &  0.739 \\
    PESQ                 &  1.427 &  1.437 &  1.470 & 1.572 &  \textbf{1.621} &  1.617 \\
    \bottomrule[1.5pt]
    \end{tabular}
\end{table}

\begin{figure}
    \centering
    \includegraphics[height=6cm]{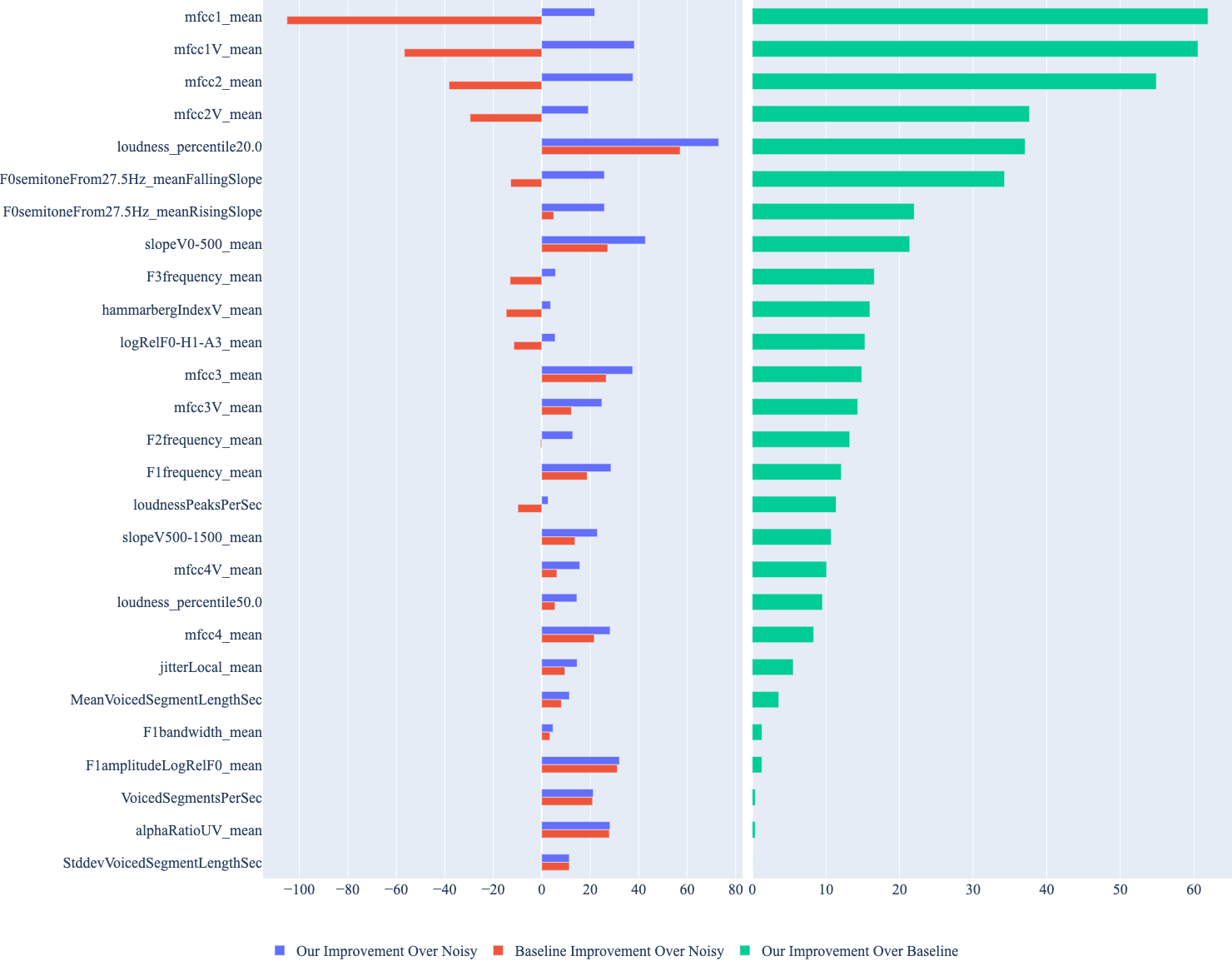}
    \caption{Acoustic improvements of the finetuned Demucs. The left side depicts the Demucs improvement over noisy in red versus the finetuned improvement over noisy in purple. The right side shows the relative improvement of our finetuned Demucs model over the baseline model}
    \label{fig:demucs_improv}
\end{figure}

\begin{figure}
    \centering
    \includegraphics[height=6cm]{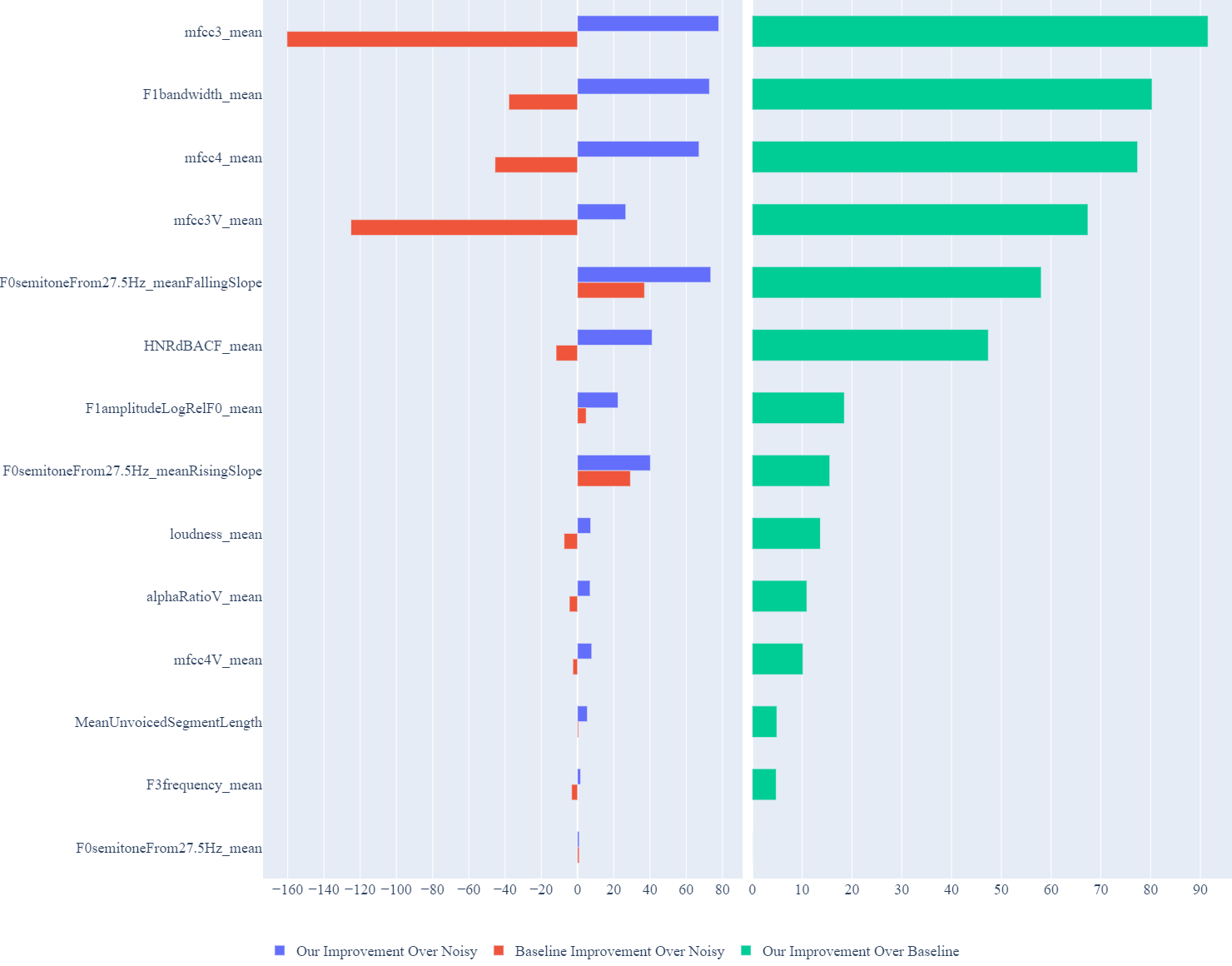}
    \caption{Acoustic improvements of the finetuned FullSubNet Model. The left side shows FullSubNet improvement over noisy in red versus the finetuned improvement over noisy in purple. The right side shows the relative improvement of our finetuned FullSubNet model over the baseline model}
    \label{fig:fsn_improv}
\end{figure}

\section{Results}

\subsection{Acoustic Improvement}
To quantify the improvement of the acoustic parameters mentioned in section \ref{eval-metrics}, we measure acoustic improvement as defined by \cite{taploss}. First, we calculate the mean absolute error (MAE) across the time axis. The MAE between our novel dataset's noisy and clean audio files is denoted as $MAE_{N}$. The MAE between the baseline enhanced audio (output of FullSubNet or Demucs without finetuning) and clean audio files are denoted as $MAE_B$. Lastly, the MAE between the enhanced audio from the finetuned model and clean audio files is denoted as $MAE_F$. Then, improvement is defined as follows, where $I_B$ is the improvement of baseline models and $I_F$ is the improvement of our fine-tuned models.

\begin{equation} \label{eq:1}
I_B= 1-\frac{MAE_{B}}{MAE_{N}}
\end{equation}
\begin{equation} \label{eq:2}
I_F= 1-\frac{MAE_{F}}{MAE_{N}}
\end{equation}

Figure \ref{fig:demucs_improv} and  \ref{fig:fsn_improv} compare the acoustic improvements for our finetuned Demucs and FullSubNet models to the baseline models. We find that our finetuned models can improve our baselines for over 14 acoustic parameters.

\subsection{Objective Metric Results}
Our results for the objective metrics are depicted in Table ~\ref{table:evaluation-table}. We find that our finetuned models are able to outperform the baseline Demucs and FullSubNet models across all metrics.

\begin{table}
  \caption{Results Table}
  \label{table:evaluation-table}
  \centering
\begin{tabular} {P{0.08 \linewidth} P{0.12 \linewidth} P{0.14 \linewidth} P{0.16 \linewidth} P{0.18 \linewidth}}
\toprule[1.5pt] \toprule[1.5pt]
{} &  \multicolumn{2}{c}{Demucs} & \multicolumn{2}{c}{FullSubNet} \\
\midrule[1pt]
{} &  Baseline &  Finetuned & Baseline & Finetuned \\
\toprule[1.2pt]
$PESQ$ &            1.727 &            \textbf{1.921} &                1.694 &                 \textbf{1.822} \\
$LLR$ &            1.432 &            \textbf{1.645} &                1.630 &                 \textbf{1.833} \\
$STOI$ &            0.864 &            \textbf{0.883} &                0.841 &                 \textbf{0.860} \\
$CSII_{high}$ & 0.665 & \textbf{0.667} & 0.652 & \textbf{0.653} \\
$CSII_{mid}$ & 0.535 & \textbf{0.572} & 0.514 & \textbf{0.539} \\
$CSII_{low}$ &            0.327 &            \textbf{0.333} &                0.299 &                 \textbf{0.334} \\
$NCM$ & 0.692 &  \textbf{0.756} &  0.650  & \textbf{0.675} \\
\bottomrule[1.5pt]
\end{tabular}
\end{table}

\section{Conclusion} In this work, we surpass the industry-standard performance and SOTA baseline architectures -- identify transmission noise as a missing component of current speech enhancement research. We achieve top performance on the VoIP DNS Challenge, improving both the transmission and background noise of audio recordings on the Google Meet To Phone track. We set a new benchmark for speech enhancement by evaluating baseline Demucs and FullSubNet models on our novel dataset. Further, we demonstrate that introducing TAPLoss into the training process and finetuning these models can further improve performance. In the future, we aim to increase our training data from 400 samples to 1200 samples in order to achieve even better performance on our models. We believe that our work can find applications in the telecom industry and directly with mobile phone manufacturers.


\section{Acknowledgements}

We would like to thank Carnegie Mellon University, Professor Bhiksha Raj, and our mentors Joseph Konan and Ojas Bhargave for their staunch support, encouragement, and guidance throughout this project.

\bibliographystyle{IEEEtran}

\bibliography{template}

\appendix

\section{Appendix}
\subsection{Calculating TAPLoss}
\label{appendix-taploss}
To calculate the TAPLoss, we define the Temporal Acoustic Parameter Estimator $A_y$. $A_y(t, p)$ represents the acoustic parameter $p$ at a discrete time frame $t$. To predict $A_y$, we define the estimator:
$$\hat{A_y} = \Tc\Ac\Pc(y)$$

The $\Tc\Ac\Pc$ function takes in a signal input $y$. It calculates a complex spectrogram $Y$ with $F = 257$ frequency bins. It then passes this complex spectrogram to a recurrent neural network to output the temporal acoustic parameter $\hat{A_y}$. TAP loss is then defined as the mean average error between the actual and the predicted estimate.

$$MAE(A_y, \hat{A_y}) = \frac{1}{TP} \sum_{t=0}^{T-1} \sum_{p=0}^{P-1} | A_y(t, p) - A_{\hat{y}}(t, p) | \in R$$

During training, $\Tc\Ac\Pc$ parameters learn to minimize the divergence of $MAE(A_s, \hat{A_s})$ using Adam optimization. Since this loss is end-to-end differentiable and takes only waveform as input, it enables acoustic optimization of any speech model and task with clean references. 

\subsection{Metrics Definitions}
\label{appendix-metrics}
\begin{table}[h]
  \caption{Evaluation Metrics}
  \label{table:metrics-table}
  \centering
  \begin{tabular}{p{0.7cm} p{6.8cm}}
    \toprule[1.5pt] \toprule[1.5pt]
    Metric & Description  \\
    \midrule[1.2pt]

    PESQ & Captures the human subjective rating of speech quality degradation caused by network conditions including analog connections and packet loss \cite{rix2001perceptual}\\
    \midrule[1pt]
    LLR & Assesses speech quality by determining the goodness of fit between clean and enhanced speech \cite{crochiere1980interpretation}
    \\
    
    \midrule[1pt]
    
    STOI & Measures the intelligibility of degraded speech signals caused by factors such as additive noise \cite{taal2010short} \\ 
    \midrule[1pt]
    CSII & Determines speech intelligibility under bandwidth reduction and additive noise conditions by measuring the signal-to-noise ratio in separate frequency bands. It is calculated in three regions: high, mid, and low, where high measures segments at or above the root-mean-square (RMS) decibel level of the speech, mid measures segments in the range (RMS-10, RMS], and low in the range (RMS-30, RMS-10] \cite{kates2005coherence}\\
    \midrule[1pt]
    NCM & Estimates speech intelligibility based on the covariance between the envelopes of the clean and noisy speech signals \cite{santos2013}\\
    
    \bottomrule[1.5pt]
  \end{tabular}
\end{table}

\begin{figure*}[h]
  \centering
  \includegraphics[width=\linewidth,height=4cm]{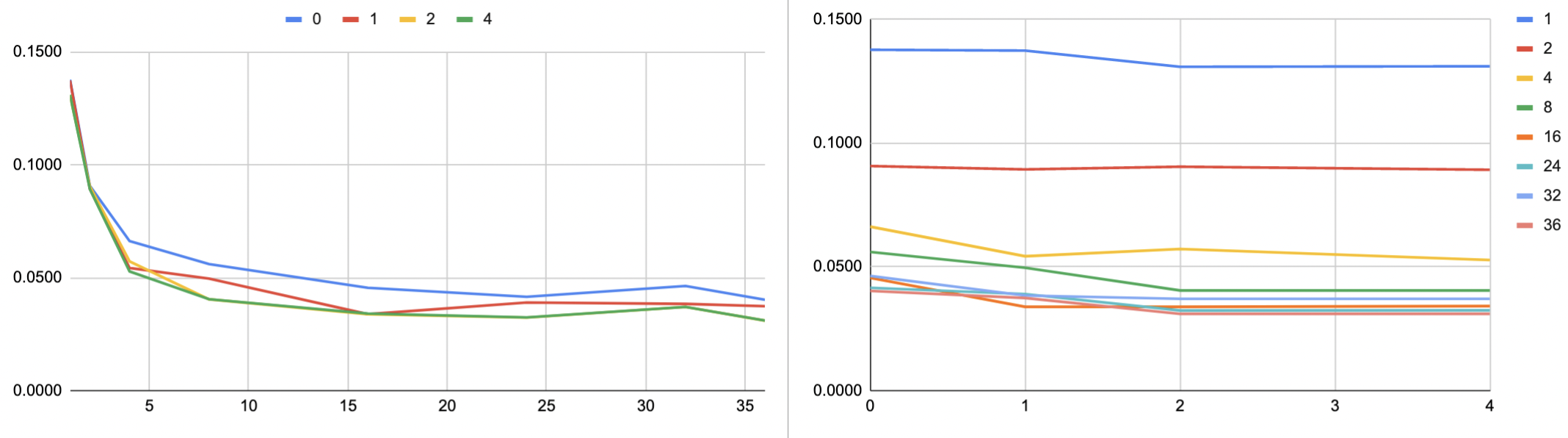}
  \caption{Demucs: The plot on the left shows the time taken per batch size for each worker. The plot on the right shows the time taken per number of workers for each batch size. This allows us to identify the optimal Demucs batch size.}
  \label{fig:train2}
\end{figure*}

\begin{figure*}[h]
  \centering
  \includegraphics[width=\linewidth,height=4cm]{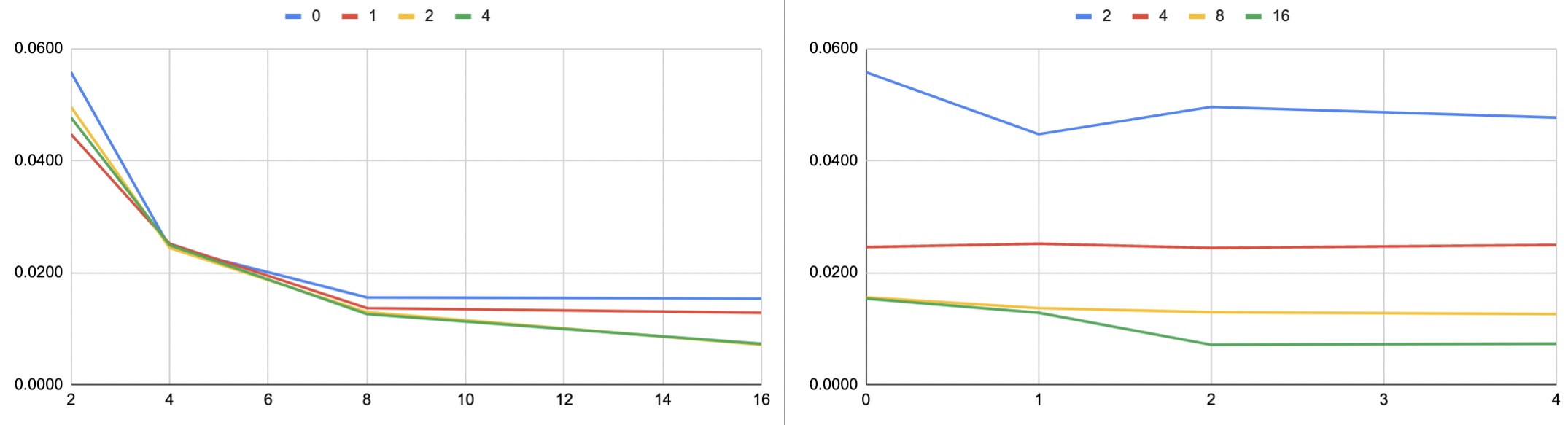}
  \caption{FullSubNet: The lot on the left shows the time taken per batch size for each worker. The plot on the right shows the time taken per number of workers for each batch size. This allows us to identify the optimal FullSubNet batch size.}
  \label{fig:train3}
\end{figure*}

\begin{figure}[]
  \centering
  \includegraphics[width=\linewidth,height=4cm]{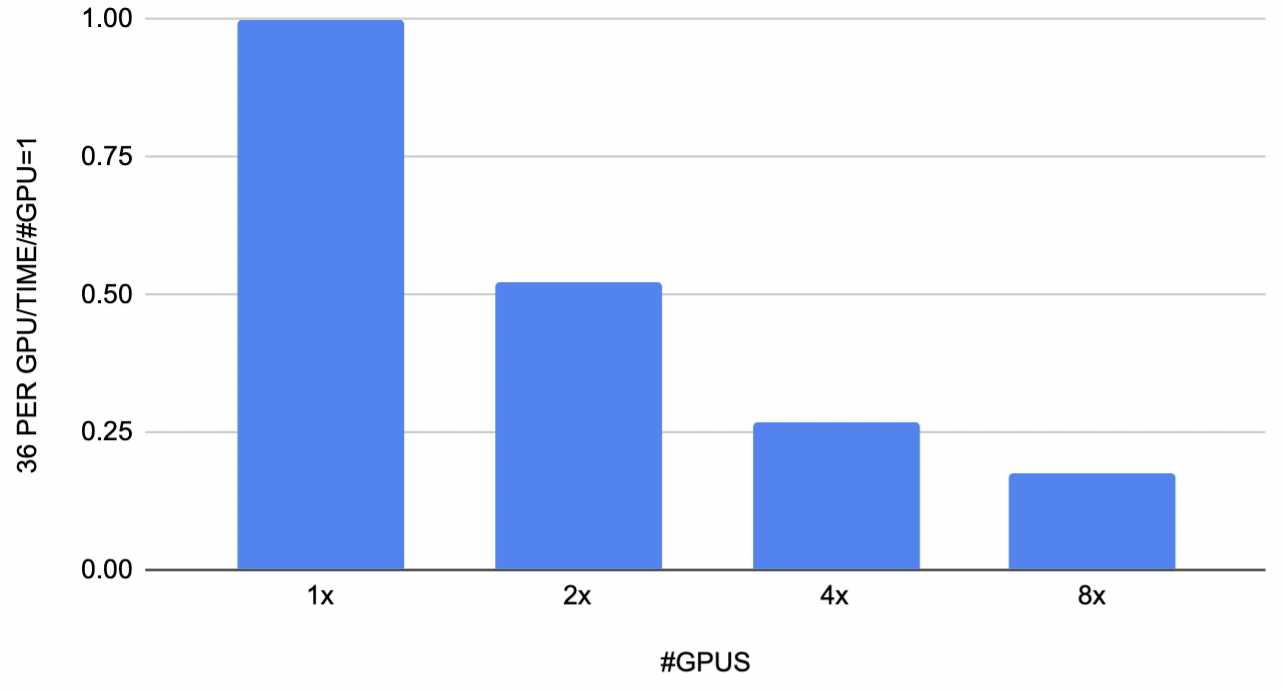}
  \caption{GPU Scaling for DEMUCS. This figure demonstrates near-perfect scaling, with a doubling of compute resulting in a doubling of throughput. However, going from 4X to 8X GPUs has diminishing returns.}
  \label{fig:train4}
\end{figure}

\begin{figure}[]
  \centering
  \includegraphics[width=\linewidth,height=4cm]{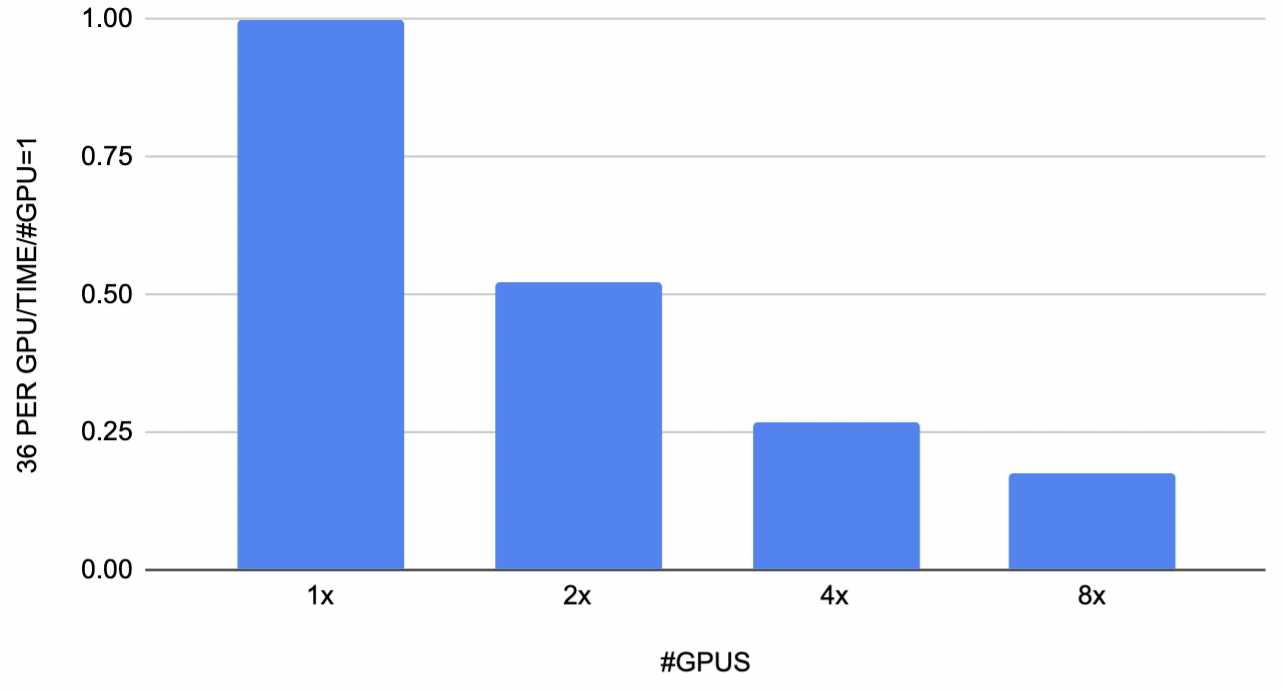}
  \caption{GPU Scaling for FullSubNet. This figure also demonstrates near-perfect scaling, with a doubling of compute resulting in a doubling of throughput. Similar to Demucs, going from 4X to 8X GPUs has diminishing returns.}
  \label{fig:train5}
\end{figure}

\subsection{Optimal Batch Size and the Number of Workers}
Figures 6 and 7 show our experiments on finding the optimal batch size for training DEMUCS and FullSubNet respectively. We finally decided on a batch size of 36 and a number of workers of 2 for DEMUCS. For FullSubNet, we went with a batch size of 16 and a number of workers of 2.

\subsection{GPU Scaling}
Once the batch size and number of workers were decided, we scaled our experiments over multiple GPUs to attain the most optimal time to be used for training as shown in figures 8 and 9.


\end{document}